\documentclass[a4paper]{jpconf}
\usepackage{graphicx}
\usepackage{wrapfig}
\usepackage{aas_macros}
\usepackage{amsmath}
\usepackage{amsfonts}
\usepackage{amssymb}

\usepackage{color}

\begin{document}
\title{Coupled MHD-Focused Transport Simulations for Modeling Solar Particle Events}

\author{Jon A. Linker$^1$,  Ronald M. Caplan$^1$, Nathan Schwadron$^2$, 
Matthew Gorby$^2$, Cooper Downs$^1$, Tibor Torok$^1$, Roberto Lionello$^1$,
and Janvier Wijaya$^1$}

\address{$^1$Predictive Science Inc., San Diego, CA 92121}
\address{$^2$University of New Hampshire, Durham, NH 03824}

\ead{linkerj@predsci.com,caplanr@predsci.com,nschwadron@guero.sr.unh.edu,
Matthew.Gorby@unh.edu,cdowns@predsci.com,tibor@predsci.com,lionel@predsci.com,
janvier.wijaya@gmail.com}

\begin{abstract}
We describe the initial version of the Solar Particle Event (SPE) Threat Assessment Tool or STAT.  STAT relies on elements of Corona-Heliosphere (CORHEL) and the Earth-Moon-Mars Radiation Environment Module (EMMREM), and allows users to investigate coronal mass ejection (CME) driven SPEs using coupled magnetohydrodynamic (MHD) and focused transport solutions.  At the present time STAT focuses on modeling solar energetic particle (SEP) acceleration in and transport from the low corona, where the highest energy SEP events are generated.  We illustrate STAT's capabilities with a model of the July 14, 2000 ``Bastille Day'' event, including innovative diagnostics for understanding the three-dimensional distribution of particle fluxes and their relation to the structure of the underlying CME driver.  A preliminary comparison with NOAA GOES measurements is shown.
\end{abstract}

\section{Introduction}
Solar Particle Events (SPEs) represent a significant hazard for humans and technological infrastructure \cite{schrijveretal2015}.  SPEs can harm aircraft avionics, communication and navigation systems.  They also represent a possible risk to the health of airline crews and passengers on polar flights.  In space, SPEs can be hazardous for crews of Low Earth Orbit spacecraft and the International Space Station, especially when engaged in extravehicular activity.  They may also imperil crews of future manned lunar or interplanetary missions.
\begin{figure}[h]
 \centering
 \resizebox{0.95\textwidth}{!}{
 \includegraphics{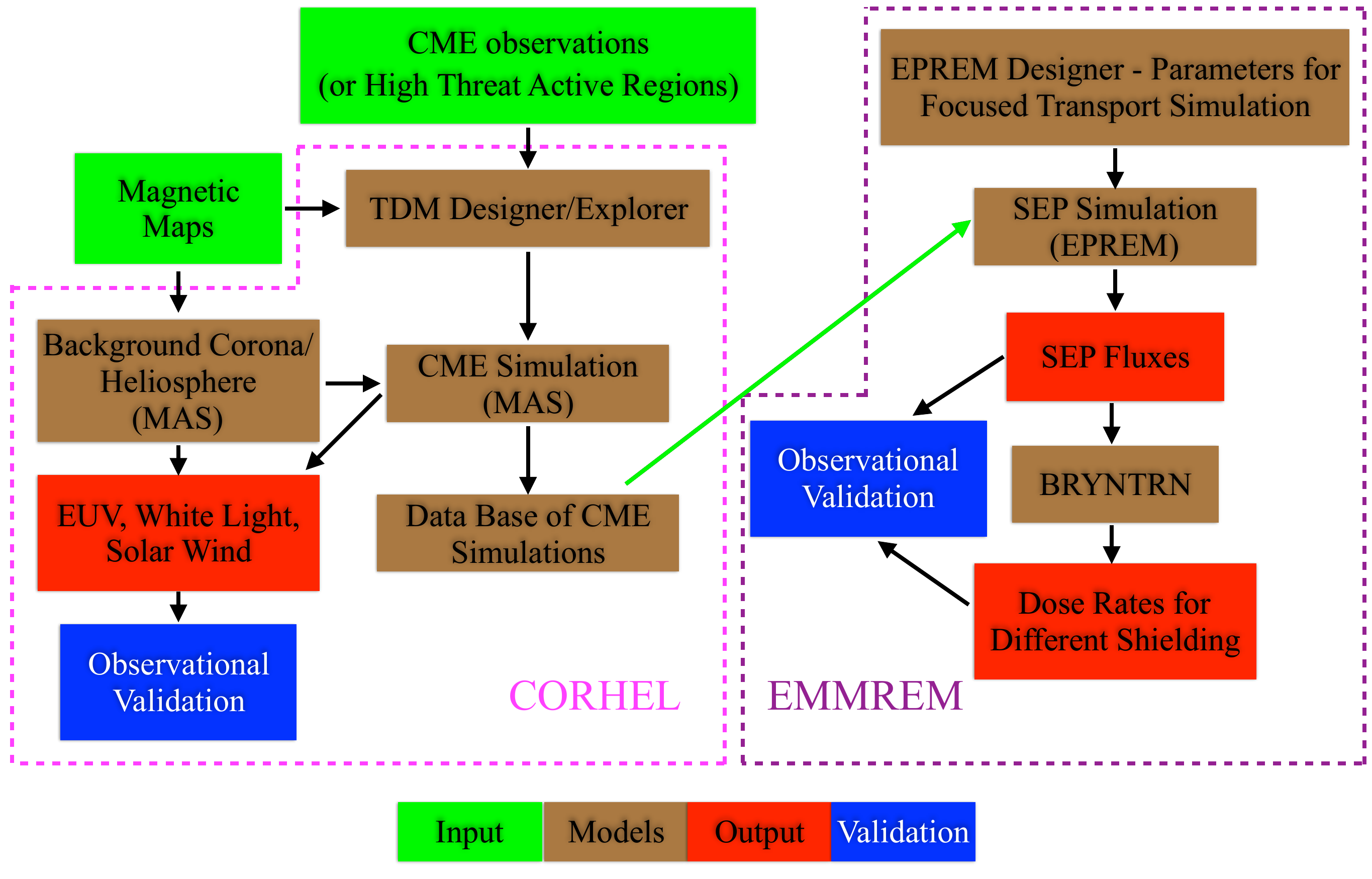}}
 \caption{Flow chart depicting the STAT framework.  The components derived from CORHEL and EMMREM are shown.
 } \label{STAT_comp}
 \end{figure}

Understanding the genesis and temporal development of SPEs spans vast regions of space and very different physical regimes.
The largest SPEs are typically associated with X-class flares and very fast ($> 1000$~km/s) coronal mass ejections (CMEs).  These events usually originate in solar active regions that contain strong, highly energized magnetic fields.  The violent release of this stored magnetic energy drives the eruption.  The solar energetic particles (SEPs) accelerated close to the Sun can travel near the speed of light, and arrive at Earth minutes to hours after the eruptive event, well ahead of the CME.  SEP generation continues as the CME propagates outward.

The exact role of flares and CMEs in particle acceleration is presently under debate; however, the compressions and shock waves associated with CMEs have long been recognized as a key driver of SEP acceleration.  The compressional properties in turn depend in part on the characteristics of the plasma through which the CME propagates.  While much can be learned from theories/models that use simplified treatments of the local acceleration environment and focus on specific aspects of the physics, the complexity of SEP observations have driven increasingly more sophisticated attempts to represent the variation in the coronal/heliospheric environment more realistically. One approach has been to use the magnetic field and plasma properties of CMEs simulated with magnetohydrodynamic (MHD) models to calculate particle acceleration and transport on selected field lines \cite{sokolovetal2004}. Considerable work has been done on the acceleration and transport of SEPs at shocks beyond 20 Rs \cite{lietal2003,li_zank2005,verkhoglyadovaetal2009} including multi-dimensional approaches 
\cite{huetal2017,zhangetal2009}.  Reference \cite{luhmannetal2017} has used the Enlil MHD model \cite{odstrcil2003} to study field line connections for multiple events in the inner heliosphere.  
%These simulations were limited to regions beyond 20 solar radii ($R_S$).

Modeling SEP acceleration and transport from the lower corona presents some of the greatest challenges,  as this is where the highest energy particles originate, and the local plasma environment has the greatest variability.  To investigate this complex but critical problem, we couple MHD simulations of CMEs in the low corona with 3D solutions of the focused transport equation for SEPs.   
Our simulations employ CORHEL (Corona-Heliosphere, \cite{rileyetal2012}) and EMMREM (Earth-Moon-Mars Radiation Environment Module, \cite{schwadronetal2010}), and have been used to model SEPs generated by a CME initiated in a generic coronal configuration \cite{schwadronetal2014}, which has led to insights into the possible origin of some of the observed properties of SEPs \cite{schwadronetal2015}.  Recently, further development of this capability has culminated in the creation of STAT (SPE Threat Assessment Tool) to model real events.  STAT allows users to run the Energetic Particle Radiation Environment Module (EPREM, a component of EMMREM)  for precomputed Magnetohydrodynamic Algorithm outside a Sphere (MAS, a component of CORHEL) simulations of real CME events to simulate SEP events and provide diagnostics that can be compared with observations.  It has recently been delivered to the Community Coordinated Modeling Center (CCMC) at NASA Goddard Space Flight Center (GSFC) for eventual runs on demand by the community.  

Figure~\ref{STAT_comp} summarizes STAT.  The present version allows users to perform coupled CME-SEP simulations for previously computed MHD simulations of CME events.  The MHD simulations have been performed with MAS/CORHEL and are stored in a database.  These are full thermodynamic MHD simulations of CMEs including the solar wind, which is the level of description required for realistic modeling of SEP events. STAT has been delivered with three CME event simulations and more will be added to the database as they become available. The user selects the parameters for a coupled CME-SEP simulation with EPREM designer; this includes choosing the CME event simulation and parameters for the solution of the focused transport equation.  At the completion of the run, automated visualizations and diagnostics (described further in section \ref{sec_diag}) are provided.  TDm Designer/Explorer allows users to create stable, pre-event flux-rope magnetic configurations using modified Titov-Demoul\'in flux ropes \cite{titovetal2014}; these are computed with magnetogram-derived boundary conditions for user-selected time periods.  At the present time, CME simulations with TDm designer use simplified physics (plasma pressure = 0, no energy equation).  In the future, it will allow CCMC users to perform thermodynamic CMEs simulations and add these to the database.

In this paper, we describe the elements of this new modeling capability and show an example for a specific event:  the July 14, 2000 (``Bastille Day'') flare/CME.  In the following sections we describe the MHD modeling (section \ref{sec_MHD}) and SEP modeling (section \ref{focused_transport}) capabilities, and the development of innovative diagnostic techniques to facilitate understanding of the results 
(section \ref{sec_diag}).  We summarize in section \ref{sec_summary}.

\section{MHD Modeling}
\label{sec_MHD}
\subsection{The Background Corona}
\label{sec_corona}
The properties of compressional regions such as shocks that drive SEP acceleration depend critically on the properties of the local plasma environment.  This is especially true in the coronal environment close to the Sun, where the plasma and magnetic properties can vary by orders of magnitude, which leads to significant variation in the local Alfv\'en ($V_A$) and sound speeds ($C_S$).  A realistic model of the background solar corona is therefore an essential component of any effort to model the acceleration of SEPs in this region.  Simulations of the solar corona for specific time periods using the MAS model have a long history of development and application \cite{mikiclinker96,linker99,mikic99a,lionello09,rileyetal2012,downsetal2013,titov17,linker17};  MAS is written in Fortran and achieves parallelism using the Message Passing Interface (MPI).  It has been shown to scale efficiently to thousands of processors.  The primary input data to MAS/CORHEL are synoptic maps of the photospheric magnetic field from observatories such as the National Solar Observatories (NSO) SOLIS and GONG, the Michelson Doppler Imager (MDI) aboard the  the Solar and Heliospheric Observatory (SOHO), and the Helioseismic and Magnetic Imager (HMI) aboard the Solar Dynamics Observatory (SDO).  The model can also ingest maps from surface flux transport models, e.g., \cite{schrijver_derosa2003} and \cite{argeetal2010}.
In the ``thermodynamic'' MHD model, a realistic energy equation that accounts for anisotropic thermal conduction, radiative losses, and coronal heating is included.  This allows the plasma density and temperature to be computed with sufficient accuracy to simulate EUV and X-ray emission observed from space.  This accuracy is vital for obtaining realistic $V_A$ and $C_S$, but greatly increases the computational requirements.   Reference \cite{toroketal2018} describes the equations solved; 66 million grid points were used in the calculation described in this paper.  A semi-implicit algorithm is used to allow for time-steps larger than the fast magneto-sonic wave limit \cite{lionello99,caplan17}.  This feature is crucial for efficient computation of the detailed structure of solar active regions  (where $V_A$ is very large) within a global model.  Recently, a wave-turbulence-driven (WTD) methodology \cite{lionelloetal2014,downsetal2016} has been applied to coronal heating in the model; this was used to predict the structure of the solar corona prior to the August 21, 2017 total solar eclipse \cite{mikicetal2018}.  

An important aspect of the modeling approach for CORHEL is to divide the coronal and heliospheric domains, and to allow for different codes and approximations in each domain.  Coronal solutions are computed from the solar surface to beyond the magnetosonic critical points (usually $20$-$30 R_S$) and are used  to provide the inner boundary condition for heliospheric solutions \cite{lionelloetal2013}.

\subsection{Modeling CMEs}
\begin{figure}[h]
 \centering
 \resizebox{0.9\textwidth}{!}{
 \includegraphics{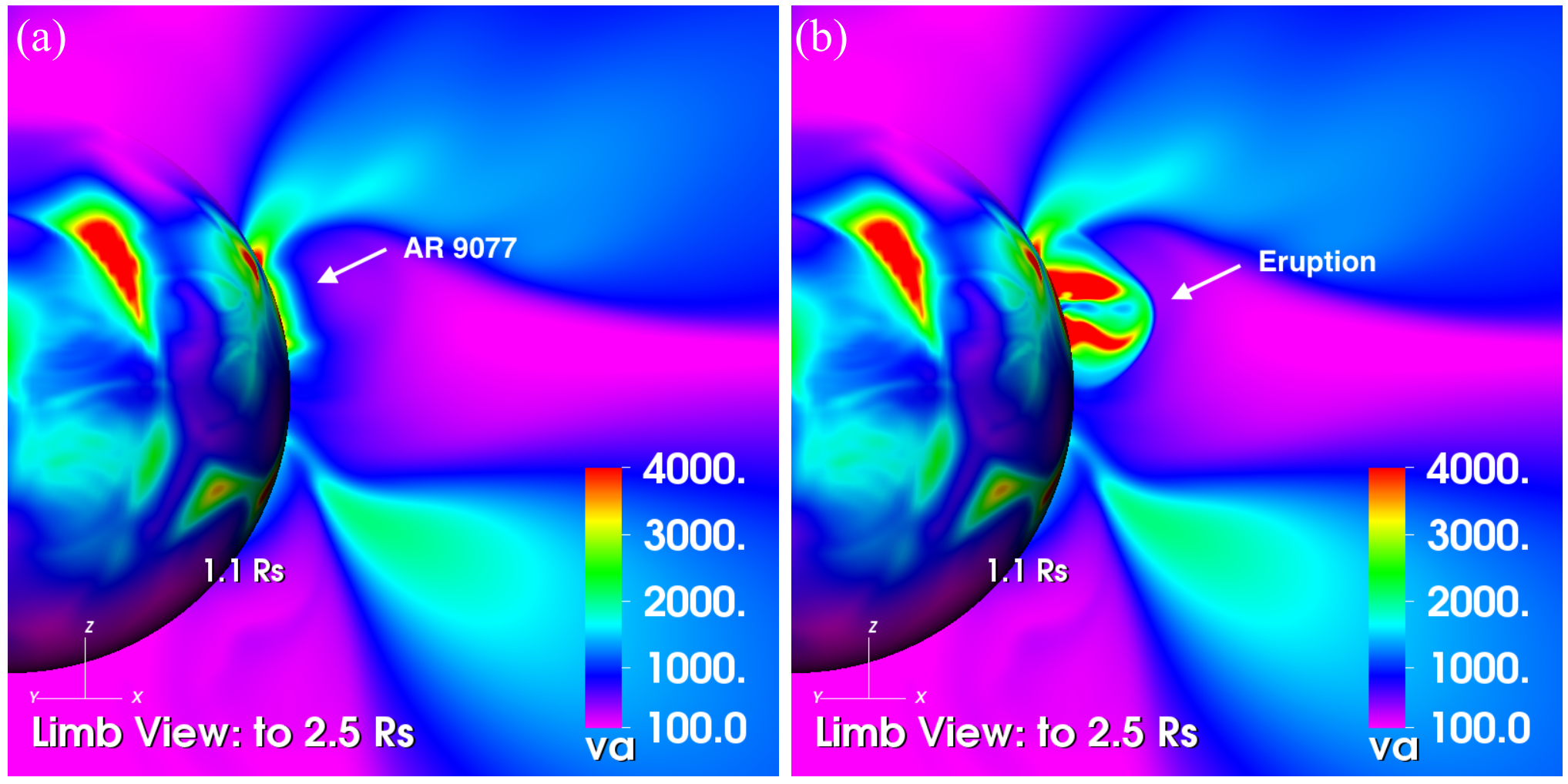}}
 \caption{Three-dimensional (3D) view of the Alfv\'en speed ($V_A$) in the simulated corona (km/s) for the Bastille Day event, contoured in color on a sphere at 1.1 $R_S$ and in a meridional plane cutting through the center of the active region.   (a) $V_A$ just prior to eruption.  (b) 2.4 minutes after the start of the eruption.
 } \label{fig_VA}
 \end{figure}
The largest SEP events start in the low corona; understanding and predicting the particle fluxes of these events requires modeling the initiation and propagation of CMEs in a realistic low coronal environment.  To robustly simulate CMEs, we employ an extension to the  Titov-Demoul\'in (\cite{titov99}, hereafter referred to as TD) model, a useful analytical model of the stable pre-event structure of a CME that has found wide application \cite{roussev03b,TorokKliem2006,TorokKliem2007}.
The flux rope can be destabilized by introducing a perturbation such as the frequently observed canceling flows near the polarity inversion line (PIL, the line separating positive and negative polarities).  In the original TD model, the entire magnetic field (both the flux rope field and the overlying arcade that stabilizes it) is described by the equilibrium.  To model real/observed CMEs in the corona, we include the background field based on observed photospheric magnetic fields.  To produce an equilibrium configuration that satisfies this, the improved model accounts for the pre-existing coronal magnetic field as well as the magnetic field created by the flux rope current itself.  Specifically, the flux rope parameters are chosen to account for the stabilizing tension of this pre-existing field.  We refer to this as the TDm or modified TD model.  With this technique, we introduce a flux rope that is close to, but beneath, the threshold for eruption \cite{lionelloetal2013,titovetal2014}.

To simulate CMEs in global coronal models with TDm,  we first separately develop a pre-event flux-rope configuration using a zero-beta MHD model (solution of the momentum equation and Faraday's law with plasma pressure = 0, an option in MAS).  These solutions can be computed rapidly (a few minutes) on supercomputers.  This relaxed configuration is then inserted into the global coronal model (a full thermodynamic MHD solution with solar wind) and relaxed further, prior to initiating the eruption.  We take care to preserve the original $B_{r0}$ (the lower boundary condition, derived from magnetogams) in developing the flux rope configuration, as this a key observational input to the model; the procedure is described by \cite{linkeretal2016} and \cite{toroketal2018}.

The July 14, 2000 CME event (often referred to as the Bastille Day event) commenced with an X5.7 flare in NOAA active region 9077.   The CME had a propagation speed of $\approx$\,1700\,km\,s$^{-1}$ in the SOHO/LASCO coronagraph field of view \cite{andrews01}. The flare was followed by an intense radiation storm that resulted in one of the 16 ground level events of cycle 23 \cite{bieber02}.  To model this event, we used a chain of seven overlapping TDm flux ropes, which together form a single rope, to account for the highly elongated and curved PIL in the source region of the eruption.  After relaxing this configuration, we introduced converging flows that cancel flux along the PIL to trigger the eruption.  
Our results, including the interplanetary propagation of the CME, are described by \cite{linkeretal2016} and \cite{toroketal2018}; here we note some aspects most relevant to modeling SEPs.

\begin{figure}[h]
 \centering
 \resizebox{0.9\textwidth}{!}{
 \includegraphics{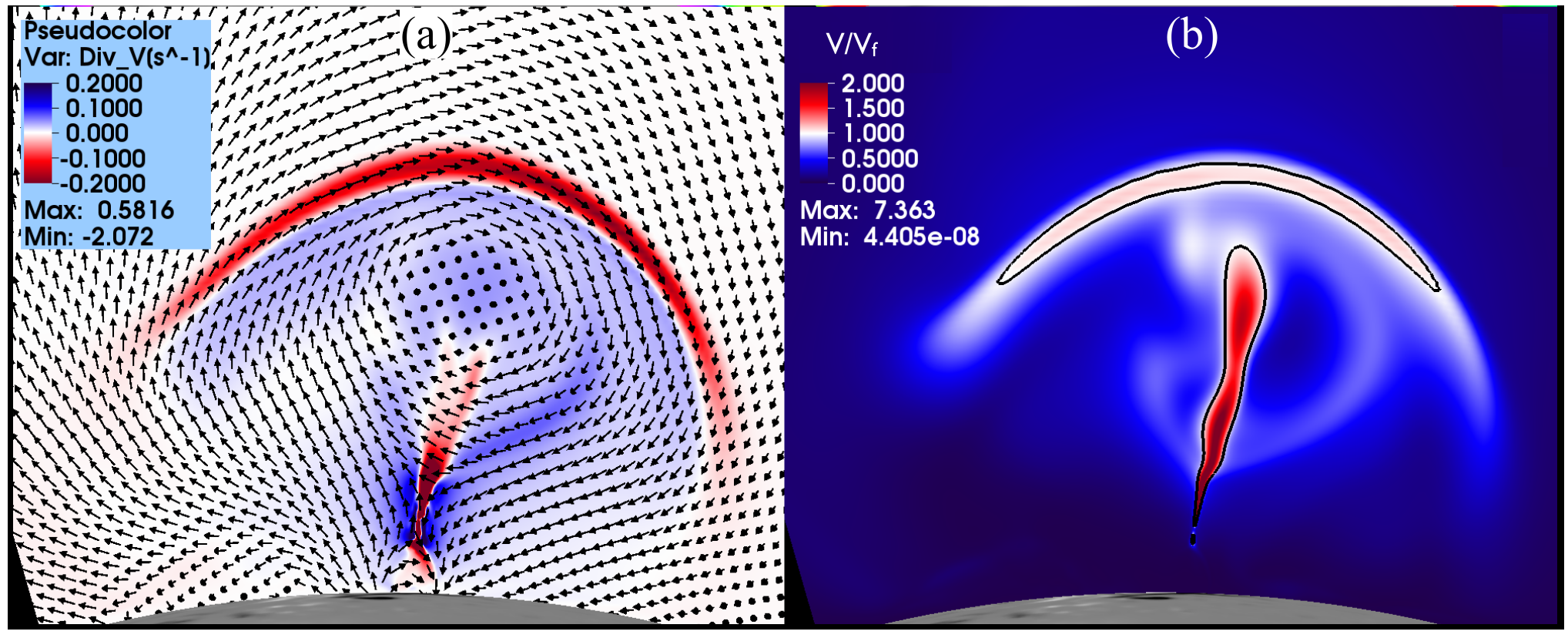}}
 \caption{Plasma properties of the simulated CME in a plane cutting through the active region, shortly after eruption.  (a) $\nabla\cdot\mathbf{V}$ (compression of the plasma) as color contours  superimposed on the magnetic field vectors projected on the plane.  (b) Local plasma speed divided by the fast most speed $v_f$ (mach number) in the same plane as (a).  
 } \label{fig_div_mach}
 \end{figure}

Figure \ref{fig_VA} shows the variable structure of $V_A$ in the simulated corona. The ``bubble'' of large $V_A$ protruding upward just after the eruption starts (labeled ``eruption'' in Figure \ref{fig_VA}b) is the erupting CME carrying strong magnetic fields upward into the corona.  The figure shows the very different local plasma properties the CME can encounter as it evolves.  Figure \ref{fig_div_mach}(a) shows magnetic field vectors low in the corona projected on a plane cutting through the active region, tracing out the approximate location of the CME flux rope soon after the eruption has started.  Superimposed on these are color contours of $\nabla\cdot\mathbf{V}$ (where $\mathbf{V}$ is velocity), with large, negative values in red (these are the regions of strongest compression).  The surface magnetic field is contoured in grey on the solar surface (barely visible in this view).  Strong compression can be seen in front of the CME, on the flanks, and beneath the flux rope where strong plasma flows associated with reconnection are jetting outward.   Figure \ref{fig_div_mach}(b) shows the magnitude of $\mathbf{V}$ divided by the maximum fast mode speed $v_f = \sqrt(V_A^2 + C_S^2)$.  

This CME simulation is in the STAT database and can be used as the basis for solutions of the focused transport equation to model particle acceleration and transport, as described in the next section.

 \section{Focused Transport Modeling of Energetic Particles}
 \label{focused_transport}
To model particle acceleration and transport in the corona and inner heliosphere, we couple time-dependent MHD solutions (like those described in the previous section) to 
EPREM, which has been designed to couple with MHD models \cite{kozarevetal2010}.  It is written in C and uses MPI for parallelism.  EPREM computes energetic particle distributions along a 3D Lagrangian grid of ``nodes'' that propagate out with the solar wind in the evolving magnetic fields of the inner heliosphere.  The details of the method of solution are described by \cite{schwadronetal2010}; here we provide a brief overview.  EPREM accounts for the time-dependent transport of pickup ions, suprathermal, and energetic particles along and across magnetic field lines using the evolving 3D field and flow topology provided by the MHD simulation.  It produces time histories of the particle velocity distribution functions as a function of pitch angle and energy.  Particle transport and energy change parallel to the interplanetary magnetic field is treated in the EPREM model with the focused transport equation \cite{ruffolo1995,skilling1971}, using the formalism outlined by \cite{kotaetal2005}: 
 \begin{eqnarray}
& \left ( 1 -\frac{\mathbf{V}\cdot\mathbf{e_b}v \mu}{c^2} \right ) \frac{df}{dt} &  \\
&+ v \mu \mathbf{e_b} \cdot \nabla f & \\
&+ \frac{\left ( 1 - \mu^2 \right )}{2} \left [ -v \mathbf{e_b} \nabla \ln B
-\frac{2}{v} \mathbf{e_b} \cdot \frac{d\mathbf{V}}{d t} +
\mu \frac{ d \ln \left (n^2 / B^3 \right )}{d t} \right ] \frac{\partial f}{\partial \mu}& \\
&+ \left [ - \frac{ \mu \mathbf{e_b} }{v} \cdot \frac{d\mathbf{V}}{d t}  
+ \mu^2 \frac { d \ln \left ( n /B \right )}{d t} 
+ \frac{ \left (1- \mu^2 \right )}{2}  \frac { d \ln \left ( B \right )}{d t}  
\right ] \frac{\partial f}{\partial \ln p } & \\
& = \frac{\partial}{\partial \mu} \left ( \frac{D_{\mu \mu}}{2} 
\frac{\partial f}{\partial \mu}\right )& 
%\\
%& - \frac{1}{p^2}  \frac{\partial}{\partial p} \left ( 
%p^2 D_{p p}  \frac{\partial f_0}{\partial p} \right ) & 
\end{eqnarray}
%where convection (1), particle streaming (2), adiabatic focusing (3), cooling (4),  pitch angle scattering (5) , and stochastic acceleration (6) are accounted for.
where convection (1), particle streaming (2), adiabatic focusing (3), cooling (4),  and pitch angle scattering (5) are accounted for.
Here $d/dt$ is the convective derivative, $\mathbf{V}$ is the solar wind flow velocity,  $\mathbf{e_b}$ is the unit vector along the magnetic field, $\mu$ is the cosine of the pitch angle, $n$ is the plasma density, $B$ is the magnetic field strength, $p$ is the particle momentum, and $v$ is the particle speed. 

The advantage of this Lagrangian formulation is that most of the transport coefficients are obtained simply by differencing the bulk plasma quantities (e.g., density, field strength, and plasma velocity) at each node moving with the solar wind flow. 
In this formulation, the grid is made up of nodes that propagate out with the solar wind flow; in the rotating frame of the Sun the nodes follow magnetic field lines.  The connected lists of node lines are referred to as streams.  This formulation enables highly efficient computation of energetic particle distribution functions at each node.  It is further described by \cite{schwadronetal2010}; an example of a node mesh is shown in Figure 4 of that paper.  EPREM also solves for cross-field diffusion and particle drift using a separate convection-diffusion equation, as described by \cite{schwadronetal2010}.  For the results shown in this paper, only parallel transport was solved for.  \cite{schwadronetal2014} describes results that did include these terms.
%In addition to solving for the parallel transport, EPREM can also solve for cross-field diffusion and particle drift with a separate convection-diffusion equation \cite{schwadronetal2010}.  
EPREM has been used to solve for pickup ion distributions \cite{schwadronetal2010,chenetal2015}, energetic particle distributions \cite{dayehetal2010,pourarsalanetal2010}, energies from keV up to relativistic GeV energies \cite{cucinottaetal2010}, and has been used to investigate SEP acceleration low in the corona using MHD solutions \cite{kozarevetal2013,schwadronetal2014,schwadronetal2015}.

In practice, the time-dependent MHD data is supplied to EPREM as a sequence of files that are read in as the EPREM calculation proceeds, and the MHD values are used to compute the terms in the above equation.  The initial background MHD solution is used to propagate the nodes of the Lagrangian grid throughout the corona and heliosphere.  As the simulation proceeds, nodes are carried out of the domain (beyond 1 AU) and are dropped from the calculation.  New nodes are added to the calculation at the lower boundary and propagate into the domain.  Closed field regions in the MHD solution (e.g. within helmet streamers) have zero or very weak flow, and are therefore not seeded with nodes.  In principle, MHD solutions for both the coronal and heliospheric domains can be included in the EPREM calculation.  At the present time we only employ the MHD coronal domain within EPREM, and fill the heliosphere using a simple spiral magnetic field created with a radially constant solar wind speed.  This restricts us to modeling the first few hours of an SEP event; once the CME leaves the coronal portion of the domain, possible SEP acceleration from the CME propagation in the heliosphere is not modeled.  Particle acceleration in the model occurs primarily through the plasma and magnetic field compressional terms in (3); therefore we expect regions in the MHD simulation undergoing strong compression to provide strong particle acceleration.  
%However, within and beneath the erupting flux rope (as in Figure \ref{fig_div_mach}), the domain is not (or very sparsely) seeded with nodes, and the region is not connected to open fields, so these regions typically don't contribute to particle acceleration in EPREM solutions.    

As part of developing STAT, we greatly optimized the EPREM computations and made the code more suitable for runs-on-demand by implementing a number of structural changes.  These included changing static arrays to dynamic allocation, using temporary stride-1 arrays, re-ordering loops, in-lining functions called within heavily-nested loops, and pre-computing expensive floating point operations outside of loops when possible.  We also found that the computation would slow dramatically in cases where the nodes ``bunch-up'' (such as in the slow flows after a CME launch, or in close proximity to the CME shock), requiring a large number of sub-cycles (for time-step stability).  In these cases the nodes were generally much closer than the grid spacing of the MHD values, making these extensive calculations superfluous.  We modified the algorithm to ignore nodes within a minimum distance (generally much less than the MHD grid size).  This modification greatly reduced the computation time while preserving the same physical solution.  As a result of optimization, low resolution runs (96 streams) that required substantial computation (10-20 hours of wall clock time on hundreds of cores) can be completed in less than an hour on a single 24-core node.  Typical resolution runs (such as shown in the next section) use 384 streams and 2000 nodes per stream and require several hours on $\sim$100 cores.  This is a small fraction of the computational requirements for detailed thermodynamic MHD simulations of CMEs.  

\subsection{Key Parameters}
Within the compressional acceleration paradigm for SEP acceleration and transport, three important parameters that influence the solutions are the parallel mean free path for scattering ($\lambda_\parallel$), the seed population of suprathermal particles ($J_{seed}$), and diffusion perpendicular to the magnetic field ($\kappa_\perp$).  These parameters are poorly constrained by observations, so at the present time we use relatively simple approximations.  Indeed, one of the purposes of creating STAT is to allow for exploration of the effects of these parameters.  We assume $\lambda_\parallel = \lambda_0(r/r_{1AU})^{2} (R_g/R_{g0})^{1/3}$, where $R_g$ is particle rigidity, $r$ is heliocentric distance, and  $R_{g0} =$ 1 GV, and  $\lambda_0 = 1$ AU.  This results in a mean free path that varies substantially over the domain.  Other scalings are available, and in principle a physical equation based on the propagation of MHD turbulence could be solved to describe its behavior \cite{sokolovetal2009}.

For the seed population we follow \cite{kozarevetal2013,dayehetal2009} and assume a differential flux that scales as
\begin{eqnarray}
J_\mathrm{seed}(T,r) = J_0 \left( \frac{r_1}{r}\right)^2 \left(\frac{T}{T_0}\right)^{-\gamma_s} \exp\left(-T/T_0\right)
\end{eqnarray}
where $T$ is kinetic energy, $T_0 = 1$ MeV, $\gamma_s = 1.96$ and
$J_0$ is given in protons cm$^{-2}$ s$^{-1}$ sr$^{-1}$ MeV$^{-1}$ (chosen to reflect the activity level of the time period being modeled).  For the results shown in this paper, we  investigated the parallel transport processes only, by neglecting particle drifts and assuming  $\kappa_\perp = 0$.  These processes will be explored in future work; \cite{schwadronetal2014} describes an initial simulation that included these effects.
%Previously we have used a fixed ratio $\kappa_\perp/\kappa_\parallel = 0.01$ \cite{schwadronetal2014}; for the results presented in the next section, the terms associated with $\kappa_\perp$ and particle drifts were turned off.

 \section{Diagnostics for Coupled CME-SEP models}
 \label{sec_diag}
The ability to globally simulate a particle event provides the opportunity to directly test whether the CME/shock paradigm can explain its major features.  To fully leverage this simulation capability, we need to obtain a global picture of the acceleration in relation to the CME properties.  This is not easy, because the simulated particle data reside on the (Lagrangian frame) nodes that are scattered within the domain.  We have developed two innovations (automated field tracing and interpolated particle fluxes) to allow automated visualizations and diagnostics to be created at the completion of EPREM calculations.

 \subsection{Visualizing CME Flux Ropes}
 \label{tracer}
A key aspect of understanding and diagnosing simulations of CMEs is obtaining visual diagnostics of the CME evolution and propagation, especially of the magnetic field evolution.  In a complex 3D simulation based on realistic magnetic fields, tracing field lines from static locations usually yields an unsatisfactory picture of the evolution.
We have found that advecting tracing points with the flow can help to capture key features of the time-dependent behavior.  In this approach, we assign footpoint positions to fluid elements, and use the flow field of the simulation to advect these fluid elements, which we then use as the location from which to trace magnetic field lines.  This approach provides powerful diagnostics for the CME flux rope, and can allow visualizations to be generated automatically, i.e., without the need for painstaking selection of representative field lines.  We have implemented these tracer points into our MAS MHD simulations, so that visualizations of the field lines in the erupting CME can be generated as part of a post-processing pipeline.  To combine this capability with the results from EPREM requires a method for visualizing the particle fluxes, as described in the next section.
\subsection{Visualization of Particle Fluxes}
\label{2D3D}
A  challenge for rapidly understanding the results from EPREM simulations is that it is difficult to identify where, in 3D space, the high energy particles are being accelerated and transported, without scanning through many different plots of stream data.  The fluxes are not on a standard mesh, but are distributed throughout the volume via the stream locations.  We require a method where we can rapidly visualize the location of energetic particle fluxes and relate these to the properties of the simulated CME (such as the location of the erupting flux rope).  In order to gain a better understanding of how the particle flux evolves in time and space, we developed a new way to visualize the particle data from EPREM.

EPREM outputs particle fluxes for a specific radius for each EPREM stream, for each energy bin.  We  define a non-uniform scale vector in r and gather the flux at each of these radii.  
%The fluxes for each radii are all stored in a single file for each stream.  We then read in the fluxes for all streams, and integrate over energy to obtain integrated flux values ($\mbox{log10}$) for the three NOAA GOES energy channels ($F_{\mbox{\tiny int}}>10, 50,$ and 100 $\mbox{MeV}$).
We integrate over energy to obtain integrated flux values ($\mbox{log10}$) for the three NOAA GOES energy channels ($F_{\mbox{\tiny int}}>10, 50,$ and 100 $\mbox{MeV}$).
We distribute the integrated fluxes onto a structured spherical grid (as described below).  For each time step of the MHD simulation, the resulting 3D logical grid of flux values defined on the grid is written out so that the fluxes can be visualized together with MHD variables.

Creating a structured 3D grid of data points from a non-structured set of points requires an interpolation technique.  We experimented with different techniques, including radial basis function interpolation (RBF) \cite{broomhead_lowe1988} and the "brute-force" Shepard \cite{shepard1968} weighted average.  We found that RBF could generate spurious maxima of flux values in sparsely sampled regions.  The Shepard technique gives acceptable results without artifacts, but was time consuming even with parallel computing ($\sim$25 minutes on a 24-core node to compute a single GOES flux bin).
To increase the efficiency, we developed a new interpolation method.  Starting with the same seed points used in the Shepard interpolation, we fill the domain with values using a region-growing methodology.  This region-growing algorithm is iterative.  For each iteration, each point in the 3D grid is checked.  If the point is not already filled with a value, it takes an average of 18 adjacent points (if there is at least one point with a value) and sets the current point to that averaged value (in a temporary array).  At the end of the iteration, the main 3D grid is set to the temporary array.  The iterations continue until all points are filled.  The results are very similar to the Shepard interpolation; the major advantage of the new region-growing method is that it  takes only 1.1 minutes to compute all three GOES energy bins on 24 cores.

\subsection{Visualization and Diagnostics from the Bastille Day Simulation}
\begin{figure}[h]
  \centering
      \includegraphics[width=0.85\textwidth]{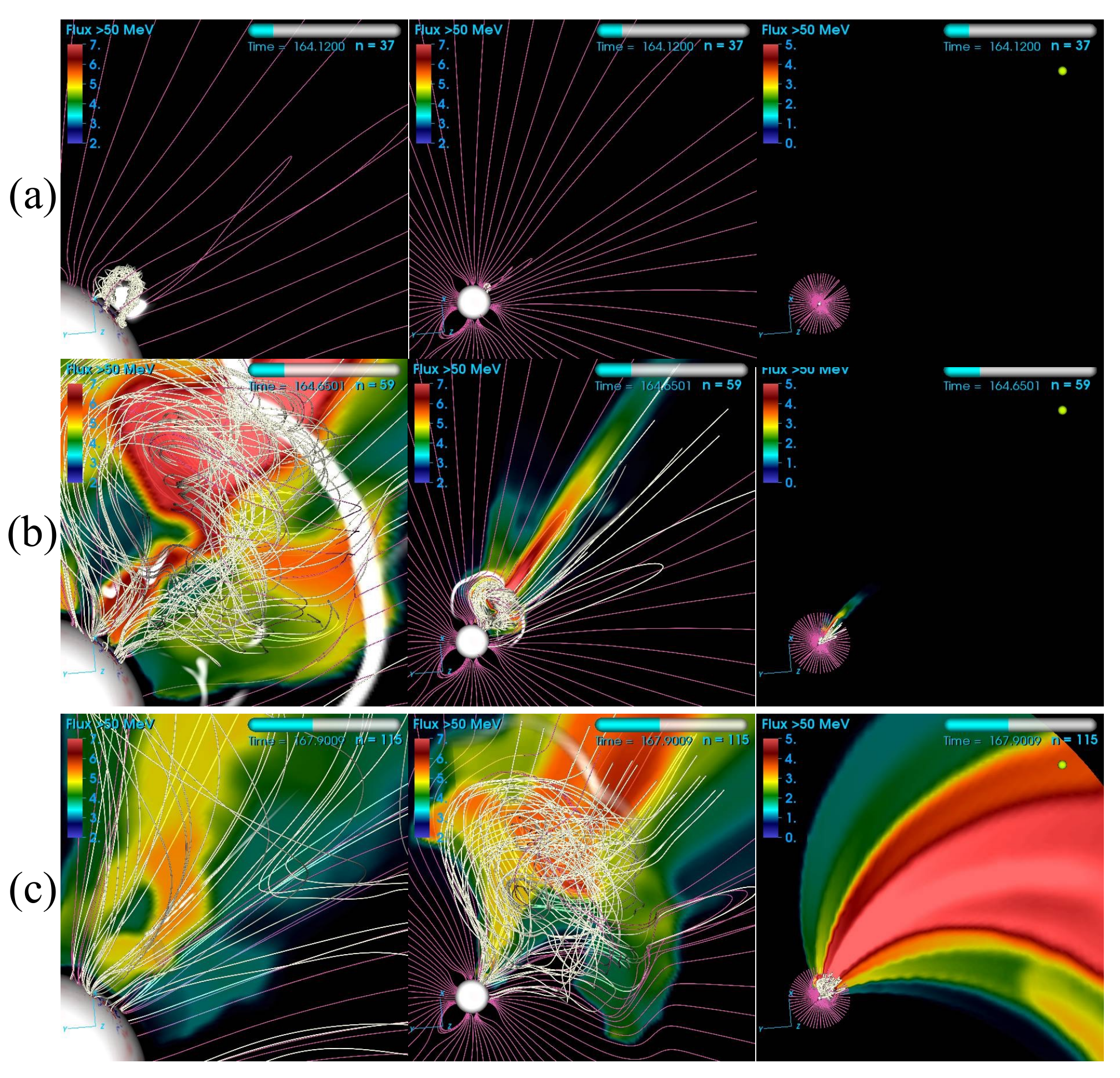}
  \caption{Combined visualization of 3D magnetic field lines, $-\nabla\cdot\mathbf{V}$ (compression), and proton flux for energies above 50 MeV (the GOES 2nd channel).  The $>50$ MeV flux is shown as color contours, and  $-\nabla\cdot\mathbf{V}$ above a threshold value is contoured in white in the heliographic plane. The figures in the leftmost column display the low coronal portion (out to about $4 R_S$) of the simulation, the middle frames display the corona to about $17 R_S$, and the rightmost frames show the domain to 1 AU (the green sphere on the righthand edge of these images shows the Earth's position.  (a) Early in the eruption phase. (b) Time = 1.8 hours in the simulation, about 14 minutes after the start of the eruption. (c) $\sim$1.5 hours after the start of the eruption.}
 \label{figureflux_fieldlines}
\end{figure} 

We illustrate the visualization techniques with results for an EPREM calculation coupled with the Bastille Day MHD simulation.  Animations of the images are part of the standard report that is provided after a STAT run is completed. 

Using tracer points for footpoints (section \ref{tracer}) and the interpolated particle fluxes (section \ref{2D3D}), we create  combined visualizations of the particle fluxes with graphical representations of the flux rope field lines; an example is shown in Figure~\ref{figureflux_fieldlines}.  Each frame shows a combined visualization of 3D magnetic field lines,   $-\nabla\cdot\mathbf{V}$ (compression), and proton flux for energies above 50 MeV (the second GOES energy channel).  The $>50$ MeV flux (created from the interpolated particle fluxes as described in section \ref{2D3D}) is shown as color contours, and  $-\nabla\cdot\mathbf{V}$ above a threshold value is contoured in white in the heliographic plane.  Each column of frames shows different portions of the domain, with the rightmost frames showing the simulation out to 1 AU.  The MHD simulation goes up to 20 $R_S$, and from 20$R_S$ to 1 AU the domain is filled with a simple spiral field as described in section \ref{focused_transport}.  At the start of the eruption (a), the CME flux rope (white field lines) is visible.  Magnetic field lines in the background corona are shown in magenta.  About 14 minutes after the start of the eruption (shown in (b)), the flux rope has propagated upward.   A driven shock (the white, ribbonlike contour) is visible in the corona, strong particle acceleration is in progress, and particles are streaming out of the corona.  One and a half hours later, shown in (c), significant fluxes are spread widely at 1 AU.  The generation of the highest particle fluxes are associated with the front of the CME low in the corona, but significant particle acceleration occurs at the flanks as well; acceleration of particles from these different regions may be responsible for energy spectra with broken power law distributions  \cite{schwadronetal2015}. 

\begin{figure}[h]
 \centering
 \resizebox{0.98\textwidth}{!}{
 \includegraphics{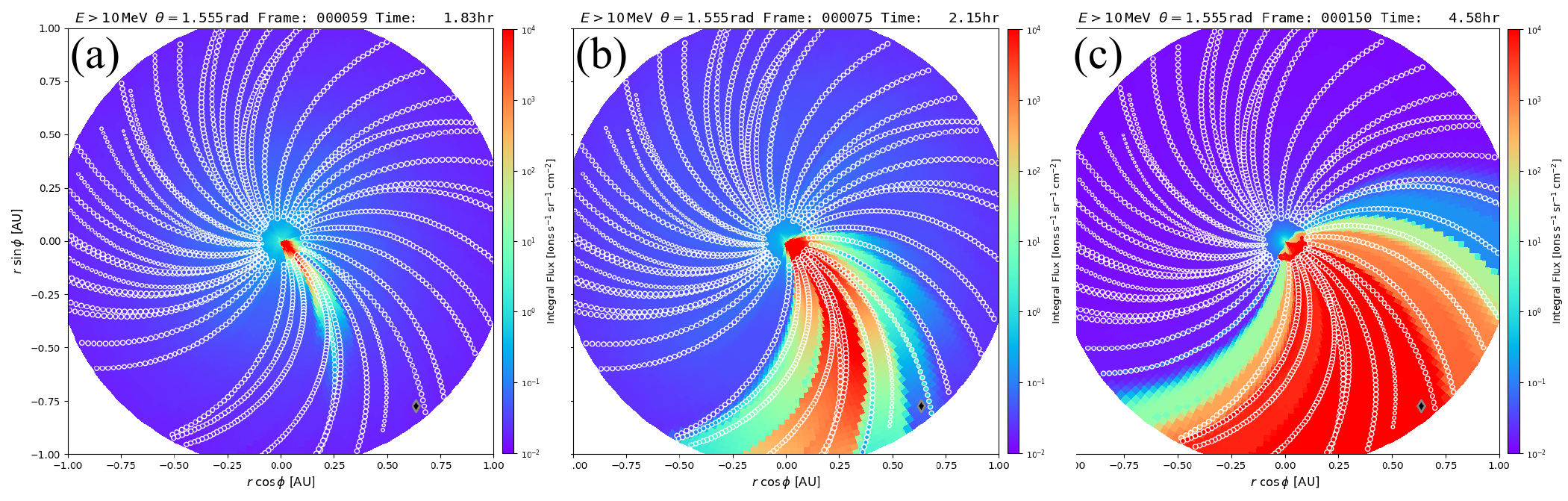}}
 \caption{2D visualization of the simulated particle flux in the first GOES energy bin ($F_{\mbox{\tiny int}}>10 \mbox{MeV}$) in the heliographic ($r{-}\phi$) plane. (a) Time = 1.8 hours in the simulation, about 14 minutes after the start of the eruption.  (b) Time = 2.15 hours, about 1/2 hour after eruption. (c) $\sim$3 hours after eruption.
 } \label{fig_2D_rphi}
 \end{figure}

Figure~\ref{fig_2D_rphi} shows the integrated particle fluxes $> 10$~MeV (the first GOES energy channel) in the heliographic plane for the Bastille Day simulation at three different times.  The black diamond shows the Earth's location.  The ``circles'' show the locations of selected EPREM nodes (where values are actually calculated) closest to this plane, and gives a measure of how well different regions are sampled to produce the images created from the interpolated fluxes.  In the simulated event, SEPs rapidly reach 1 AU; the Earth's location received significant fluxes but higher fluxes propagated eastward of the Earth.

\begin{figure}[h]
\centering
 \resizebox{0.9\textwidth}{!}{
\includegraphics{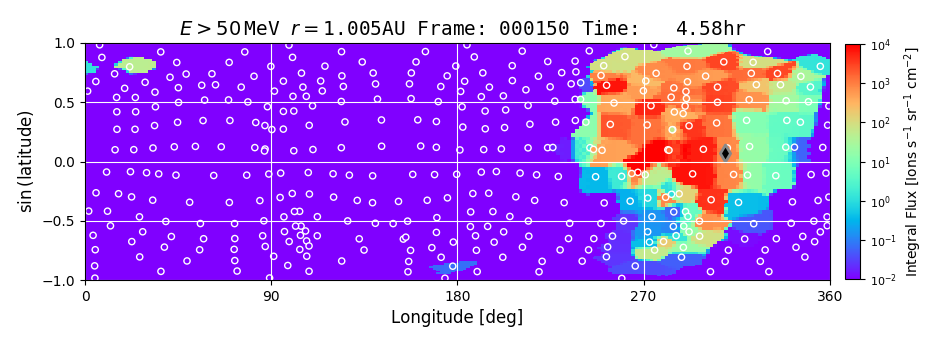}}
\caption{Simulated particle flux in the first GOES energy bin ($F_{\mbox{\tiny int}}>10 \mbox{MeV}$) at 1 AU, as a function of sin(latitude) and heliographic longitude.  The location of selected EPREM nodes for the particle calculation are shown as circles.  Earth's location is the black diamond.} 
\label{fig_sinlatlong}
\end{figure}

Figure~\ref{fig_sinlatlong} shows the particle fluxes in the first GOES energy bin at time = 4.58 hours (about 3 hours after the start of the eruption, the same time as Figure~\ref{fig_2D_rphi}(c)) on a spherical shell at 1 AU.  The shell has been stretched out as a sin(latitude)-heliographic longitude map.  Strong, but highly variable, particle fluxes are seen near Earth with significant spread in longitude and latitude.
%The EPREM nodes are distributed somewhat sparsely near Earth; in the present version of the model, the node locations are determined by the solar wind flow and are thus outside of our control.  We are presently investigating a new technique where the nodes can be seeded backward from heliospheric locations, to ensure that regions of interest are more uniformly sampled.
 
\begin{figure}[htbp]
\centering
 \resizebox{0.9\textwidth}{!}{
\includegraphics{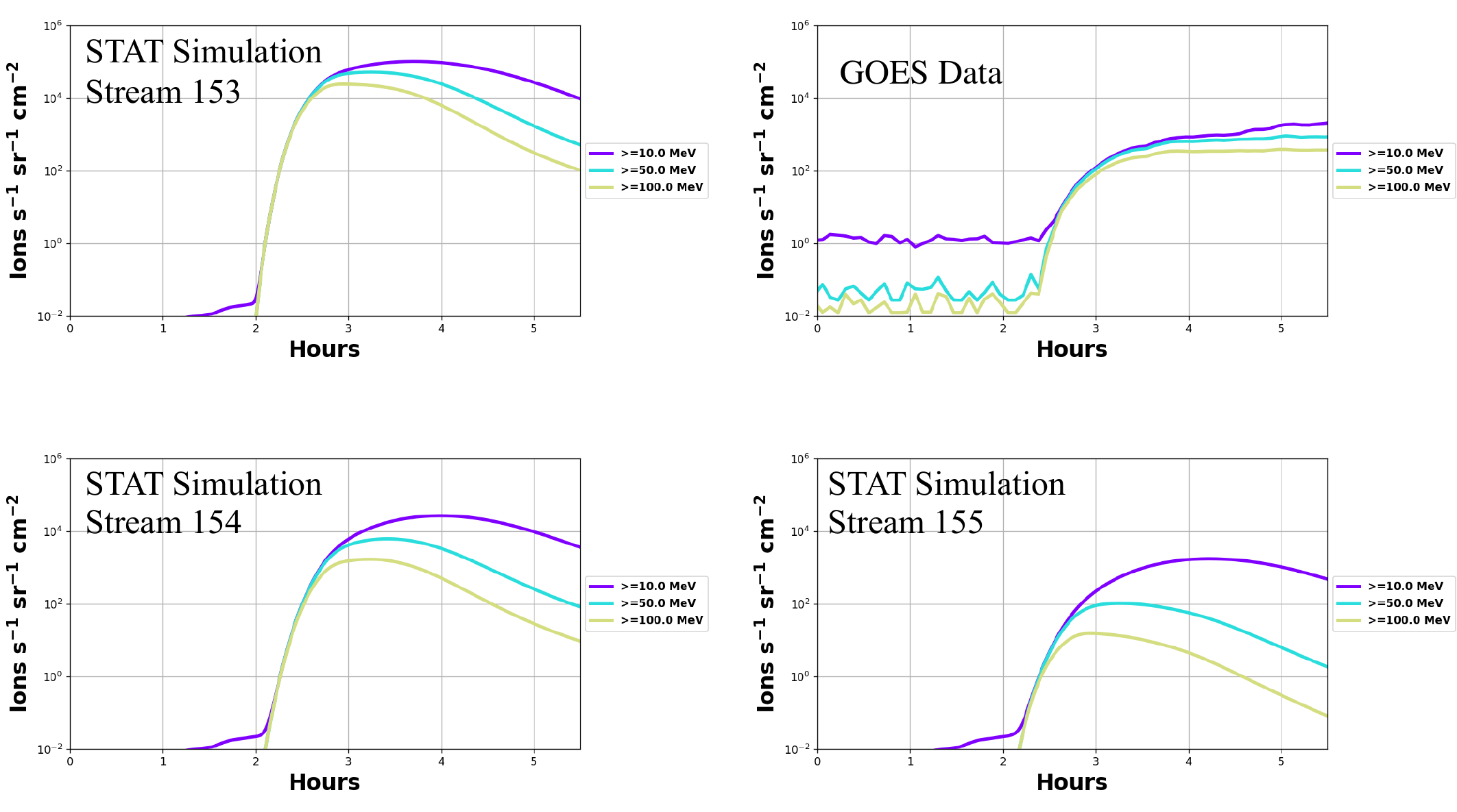}}
\caption{Comparison of simulated GOES fluxes at 3 locations near Earth with measured GOES fluxes.  The eruption starts at $\sim$1.6 hours on the time scale shown in the plots.} 
\label{fig_3_streams}
\end{figure}

To test the simulation results against observations, we provide plots and data that can be compared with GOES measurements.  To understand the particle sources, it is essential 
 to be able relate these results at a specific location to the overall 3D structure of the corona and heliosphere.  To facilitate this identification, STAT provides a clickable interface for images like Figure~\ref{fig_sinlatlong} that allows the user to browse particle fluxes at different locations.  The user can select the node locations and view the temporal evolution for the GOES energy bins, as shown in 
Figure~\ref{fig_3_streams}.  This figure shows the simulated fluxes at three of the streams close to Earth in the simulation.  Stream 154 is the circle closest to Earth (partially obscured by the black diamond) in Figure~\ref{fig_sinlatlong}; streams 153 and 155 are the adjacent circles.  Also shown in Figure~\ref{fig_3_streams} is the actual GOES data for this event.  The CME propagates out of the coronal domain of the simulation at just past 5 hours, so a comparison with GOES observations is meaningful only up to this time.  The behavior of the fluxes show similarity to the GOES observations early in the event, however, the figure illustrates the importance of investigating the variability in the solution, as the three locations give different values.  This variability might be lessened if $\kappa_\perp$ and particle drifts were included in the solution.
 
\section{Summary}
\label{sec_summary}
We have described STAT, a tool for investigating CME-driven SEP acceleration and transport in real events using solutions of the focused transport equation coupled with MHD models.  
A simulation of the July 14, 2000 CME is used to illustrate the combined model.  Full thermodynamic MHD solutions (that account for coronal heating and energy transport) are necessary obtaining realistic Alfven and sound speeds in the simulated corona.  In the simulated event, the erupting CME produces strong compressions and shocks very low in the corona.  These compressions drive significant particle acceleration in the focused transport simulations, and energetic particle fluxes stream outward and rapidly arrive at Earth orbit.  We described innovative techniques for distributing the simulated particle fluxes on a logical grid to understand their 3D structure and visualize them in relation to MHD features, such as the erupting flux rope. 
Our results show that even for a complex event like Bastille Day, this modeling approach can produce particle fluxes with characteristics reasonably similar to interplanetary measurements, indicating that this is a promising direction for tackling the difficult problem of SPEs.

We consider the results we have presented here to be preliminary, as we are still exploring how the variation of parameters can affect the solutions.  We plan to publish a more detailed study of this event when this exploration is completed.  Future planned improvements to this modeling include increasing the number of streams in the EPREM calculations, and extending EPREM to include both the coronal and heliospheric domains from the MAS simulations.  The latter extension will allow STAT to follow the entire development of an SPE that can occur over days as the CME propagates in interplanetary space.
\ack
%\section{Acknowledgments}
This research was supported by AFOSR, by NASA via the LWS TR\&T, HSR, and STTR programs, and by NSF.  Computational resources were provided by NSF's XSEDE and NASA's NAS. 

\section*{References}
\bibliographystyle{iopart-num}
\bibliography{astronum2018}

\end{document}